%% file: manuscript.tex
\documentclass[twocolumn,showpacs,preprintnumbers,amsmath,amssymb,aps,prl]{revtex4}
\usepackage{graphicx}
\begin{document}

\title{
Active Matter Transport and Jamming on Disordered Landscapes    
} 
\author{
C. Reichhardt and  C. J. Olson Reichhardt}
\affiliation{
Theoretical Division,
Los Alamos National Laboratory, Los Alamos, New Mexico 87545, USA
} 

\date{\today}
\begin{abstract}
We numerically examine the transport of active  
run-and-tumble particles driven with a drift force
over random disordered landscapes
comprised of fixed obstacles.  
For increasing run lengths, the net particle transport initially increases
before reaching a maximum and decreasing at larger run lengths. 
The transport reduction is associated with
the formation of cluster or 
living crystal states that become locally jammed or 
clogged by the obstacles. We also find that the system dynamically jams 
at lower particle densities when the run length is increased.
Our results indicate that there is an optimal activity level for 
active matter transport 
through quenched disorder, and
could be important for understanding biological transport in complex
environments or for applications of active matter particles in random media.      
\end{abstract}
\pacs{64.75.Xc,47.63.Gd,87.18.Hf}
\maketitle

\vskip2pc
There has been tremendous growth in interest 
in what are termed active matter systems,
where the individual units comprising the system
undergo self-propulsion \cite{1,2,3}.
Biological examples of such systems 
include swimming bacteria \cite{4,5}, crawling cells \cite{3}, 
and flocking or swarming particles or agents \cite{6}. 
Non-biological
active matter, such as
self-driven colloids \cite{7,8,9,10}, artificial swimmers \cite{11}, and 
assemblies of self-motile mechanical devices such as bristle bots \cite{12},
has been the focus of an increasing number of studies.
Active matter systems exhibit a rich variety of behaviors that 
disappear 
when the fluctuations experienced by the particles are 
only equilibrium or thermal in nature.  
Self-driven particles 
can exhibit
run-and-tumble dynamics similar to that obeyed by many types of
bacteria \cite{4,5}
where the particle moves in a fixed direction
for a period of time before undergoing a tumbling event and 
changing directions;
alternatively, in the case of 
active Brownian particles, 
the particles are driven by a motor in a direction that slowly rotates
due to a noise term, a dynamics
that appears in many self-driven colloidal systems 
\cite{2,7,10,13,14}. 
It was recently shown
that active Brownian particle dynamics and 
the run-and-tumble dynamics can be mapped into each other
and can be considered equivalent \cite{17}. 

When collections of repulsively interacting particles 
such as hard disks undergo thermal
fluctuations, they form a uniformly dense state; 
however, when the particles are active there can be a transition to 
a phase-separated state consisting of high density clusters residing in
a low density active gas \cite{10,13,14,15,16,17,18,19,20}. 
In two-dimensional (2D) systems of monodisperse active disks, 
the clusters have internal hexagonal or crystalline ordering \cite{10,16,18}. 
Recent experiments on light-activated self-driven colloidal particles 
that 
form such crystalline 
cluster states 
revealed 
that the clusters are highly dynamical, frequently breaking up and reforming,
so that the cluster state
has been termed a living crystal \cite{10}.
Similar living crystals have also been experimentally observed 
in other self-driven colloidal systems \cite{18}. 
The onset of the cluster phase depends both
on the density of the system and on the activity level, defined as
the distance the particles
effectively move in a fixed direction during each run, 
with cluster phases occurring at lower densities when the
activity is increased.

Studies of active matter have generally focused on systems with
a smooth substrate.
Understanding how active matter particles move in random or complex 
landscapes
is an open question that is relevant   
to biological systems which, in the wild, typically operate
in a complex environment.
Possible applications utilizing active particles
could also require the particles to interact with 
disordered landscapes.  
In studies of
active particles interacting with arrays of asymmetric barriers, 
ratchet effects arise when
the activity causes a breaking of 
detailed balance in the particle-barrier interactions 
\cite{5,21,22,23,24,25}. 
Numerical work has shown that 
flocking active particles 
can have their order enhanced by obstacle arrays \cite{26}, and that
disordered substrates can localize or trap active particles \cite{27}.

In this work we focus on the transport of active matter 
over a disordered substrate in the presence of
an external drift force. 
Particle transport over disordered substrates under an external drift  
is a very general problem that is studied in 
a variety of condensed matter systems 
such as colloids moving
over random substrates \cite{28}, 
vortices moving in dirty type-II superconductors \cite{29}, 
sliding charge density waves \cite{30},
classical charge transport \cite{31} 
and the motion of magnetic domain walls \cite{32}.
In these systems, the particle transport
in the direction of the external drift increases with
increasing thermal fluctuations, which
diminish the effectiveness of 
pinning by the substrate. 
Active particles might be expected to experience significantly reduced
pinning effects from the substrate
since, as the activity or run length of the particles is increased, 
the particles would be better able
to escape from effective trapping regions;  
however, in this
work we show that for active matter this is generally not the case.
We find that
an increase in the activity can 
initially increase the particle transport;
however, when the run lengths are large enough, the transport becomes 
strongly impeded due to local jamming effects 
caused by the formation of 
living crystals that act like rigid objects, permitting an entire cluster of
particles to be pinned by a small number of obstacles.
Our results show that there is an optimal run length or activity 
at which 
maximum transport of the active matter through disordered media
can be achieved.  We also demonstrate that the
jamming transitions observed in 2D assemblies of non-active disks
\cite{33,34,35} can be connected 
to active matter systems and that activity can form a new axis of the
jamming phase diagram.  

{\it Simulation---}   
We consider 2D systems of size $L \times L$ with periodic boundary conditions 
in the $x-$ and $y-$directions containing an 
assembly of $N_a$ active
disks with radius $r_{d}$ that interact with each other
via a harmonic repulsion. 
The dynamics of disk $i$ located at ${\bf R}_i$ is obtained by integrating
the following equation of motion:
\begin{equation}
\eta \frac{d {\bf R}_{i}}{dt} = 
{\bf F}^{m}_{i} + {\bf F}^{s}_{i} +  {\bf F}^{b}_{i} +  {\bf F}^{D}, 
\end{equation} 
where $\eta = 1.0$ is the damping constant. 
Under the motor force
${\bf F}^{m}_{i}$, a particle
moves in a fixed direction with a constant force 
$F^{m}$ for a fixed run time $\tau_{r}$, and at the end of each running
time a new running direction is chosen at random.
We select values of $F_{m}$ such that the
particles can never pass through one another or through the obstacles.  
A single particle
in the absence of other particles or obstacles 
moves a distance $R_{l} = F^{m}\tau_{r}$ in one running time. 
The steric disk-disk repulsion is modeled as
${\bf F}^{s}_i = \sum^{N_a}_{i \neq j}(2r_{d} - |{\bf r}_{ij}|)
\Theta(2r_{d} -|{\bf r}_{ij}|){\hat {\bf r}}_{ij}$,
where ${\bf r}_{ij} = {\bf R}_{i} - {\bf R}_{j}$,
${\hat {\bf r}}_{ij} = {\bf r}_{ij}/|{\bf r}_{ij}|$,
and we take $r_d=0.5$.
The term ${\bf F}^{b}_{i}$ represents the particle-obstacle interactions,
where we model the obstacles as $N_p$ immobile disks that have the same
steric interactions as the active disks. 
The disk density $\phi$ is determined by the area covered by all
the disks, $\phi=L^2/N\pi r_d^2$, where $N=N_a+N_p$; here, 
we consider systems with $L=100$.
In the absence of 
any obstacles or thermal fluctuations, 
the disks form a hexagonal solid at $\phi \approx 0.9$. 
The term ${\bf F}^{D}=F_d{\bf \hat x}$ represents the external drift force. 
In order to measure transport
we calculate the average drift velocity 
$\langle V_{x}\rangle = (1/N)\sum^{N}_{i= 1}{\bf v}_{i}\cdot { \hat {\bf x}}$.
In the absence of a drift force, 
$\langle V_{x} \rangle = 0$, and in the 
absence of any obstacles, all the particles move
with $\langle V_{x}\rangle = F_{d}$.   

\begin{figure}
\includegraphics[width=3.5in]{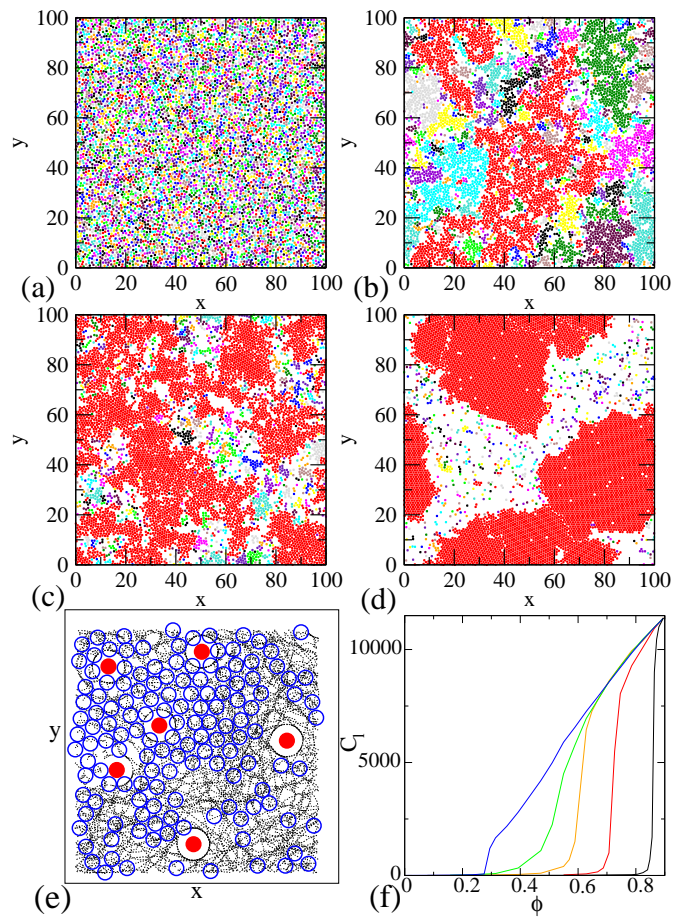}
\caption{ 
(a-d) Images of disk positions in an obstacle-free system with
$\phi=0.667$
and $N_p=0$
for increasing run lengths $R_{l}=$ (a) $0.4$, (b) $4.0$, 
(c) $20.0$, and (d) $100.0$.
Individual clusters are indicated with varying shading (coloring), with the
largest cluster marked in red.
In (d), the system has completely phase separated into 
a large cluster with local density near $\phi = 0.9$ surrounded by
a low density gas of particles. 
(e) Positions of pinned disks (filled circles), active disks (open circles),
and disk trajectories over a period of time (dashed lines) in a small
section of a sample with $N_p=200$ for $R_l=20$.
(f) $C_l$, the number of particles in the largest cluster, 
vs $\phi$ in the $N_p=0$ system for 
$R_{l} = 40$, 20, 1.4, 0.2, and $0.04$, from left to right. 
The onset of clustering occurs at higher $\phi$ 
for decreasing run length. 
}
\label{fig:1}
\end{figure}

{\it Results---}
We first characterize the onset of clustering in the absence of 
quenched disorder.
To quantify the amount of clustering present,
we utilize a cluster-identification algorithm 
described in Ref.~\cite{36}. 
In Fig.~1(a-d) we show the 
onset of cluster formation at a fixed density of $\phi = 0.667$ 
for increasing run lengths $R_{l}$. 
Different clusters are highlighted by different shadings.
At $R_{l} = 0.4$ in Fig.~1(a), we find a uniform liquid state,
while at $R_{l} = 4.0$ in Fig.~1(b), small clusters begin to form. 
The clusters grow in size and become better defined
at $R_{l} = 20.0$ in Fig.~1(c), and at large $R_{l}$ the 
system is completely phase separated into a crystalline
hexagonally ordered
solid phase with a local density of $\phi  = 0.9$ 
surrounded by a low density gas phase, as shown in Fig.~1(d) for 
$R_{l} = 100.0$. 
This behavior is identical to that found in simulations 
of active Brownian particles at a fixed density when 
the persistence length is increased \cite{16}.
Cluster formation also occurs when $\phi$ is increased for fixed $R_{l}$. 
In Fig.~1(f) we plot $C_l$, the average number of particles in the
largest cluster, versus $\phi$ for systems with 
$R_{l} = 40$, 20, 1.4, 0.2, and $0.04$.
At the smallest run length of $R_{l} = 0.04$, 
clustering does not occur until the density
is well above $\phi = 0.8$, while for  
large $R_{l}$ the onset of clustering shifts to much lower values
of $\phi$. 

\begin{figure}
\includegraphics[width=3.5in]{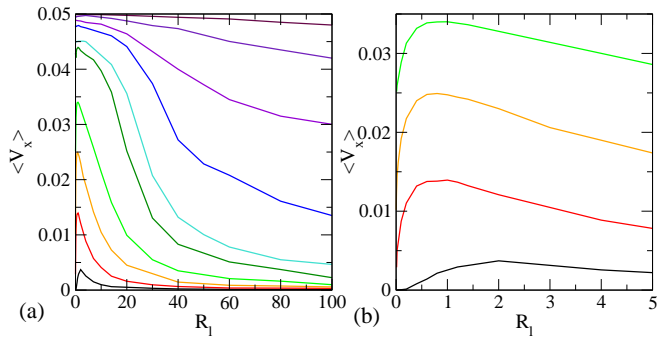}
\caption{
(a) 
$\langle V_x\rangle$, the average velocity per particle in the drift direction,
vs $R_{l}$ 
for a system with 
$\phi=0.667$
and $F_d=0.05$.
The obstacle density 
$\phi_{p} = 0.00039$, 0.00157, 0.00472, 0.0785, 0.0157, 0.02356, 0.055, 
0.0942, 0.01413, and $0.188$, from top to bottom.
When $R_{l} > 5.0$, $\langle V_{x}\rangle$ 
decreases with increasing $R_{l}$. 
(b)
$\langle V_x\rangle$ vs $R_l$ passes through a maximum for all the
curves, as shown more clearly in this blow-up of panel (a) for
$\phi_{p} = 0.055$, 0.094, 0.1413, and $0.188$, from top to bottom.       
}
\label{fig:2}
\end{figure}

We next consider the effect of adding $N_p$ obstacles in the 
form of immobile disks, as illustrated in Fig.~1(e).
We report both
the total density $\phi$ of the system which includes both the active and 
stationary disks, as well as the density 
of only the stationary disks $\phi_{p}=L^2/N_p\pi r_d^2$.
All the active disks experience
a uniform drift force with $F_{d} = 0.05$, 
so that in the absence of any obstacles $\langle V_x\rangle=F_d=0.05$
independent of the value of $R_l$.
In Fig.~2(a) we plot $\langle V_{x}\rangle$ 
versus $R_{l}$ 
at $\phi= 0.667$ for varied $\phi_{p} = 0.00039$, 0.00157, 0.00472, 0.0785, 
0.0157, 0.02356, 0.055, 0.0942, 0.1413, and $0.188$.
For $R_{l} > 5.0$ 
the particle drift velocity {\it decreases} with 
increasing $R_{l}$. 
Additionally, increasing the density of obstacles 
lowers the drift velocity.
In each case, $\langle V_x\rangle$ passes through a local maximum as a
function of $R_l$, as shown more clearly in Fig.~2(b)
where the $\langle V_{x}\rangle$ vs $R_{l}$ 
curves for $\phi_{p}  =  0.055$, 0.094, 0.1413, and $0.188$ 
peak around $R_{l}= 1.0$. 
For very short run lengths
$R_l<0.1$ for the largest obstacle density of $\phi_{p} = 0.188$, the
system jams
and $\langle V_{x}\rangle \approx 0$. 
The initial increase
in $\langle V_{x}\rangle$ with increasing $R_{l}$ 
has the same form that would appear
if increasing thermal 
fluctuations were added to the system:
$\langle V_{x}\rangle = A\exp(U/k_{B}T)$, 
where $A$ is the driving force. 
In the case of thermal fluctuations, the drift velocity would
saturate at large $T$ when the substrate loses its effectiveness.
In the active matter system for small $R_l$, increasing $R_l$ is
is similar to increasing the temperature; however, once $R_l$ is large
enough for
clustering to occur, the particle density
becomes inhomogeneous and $\langle V_x\rangle$ drops. 
We have examined a variety of values of $\phi$ and $\phi_p$
and find that in general, for $\phi > 0.4$, when quenched disorder
is present the 
drift velocity is maximized at an optimal value of $R_l$.

\begin{figure}
\includegraphics[width=3.5in]{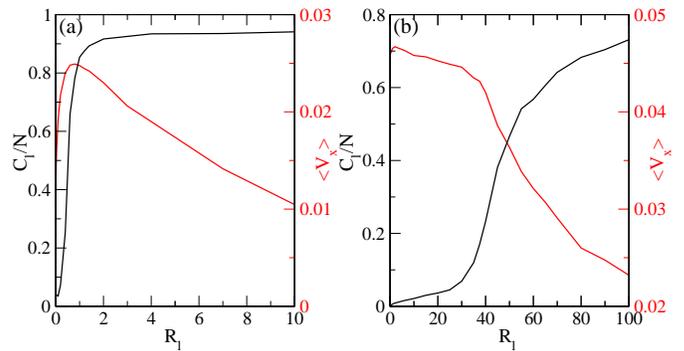}
\caption{
(a) $C_l/N$, the fraction of particles in the largest cluster (dark curve),
and $\langle V_x\rangle$ (light curve)
vs $R_{l}$ 
for a system with
$\phi = 0.667$ and $\phi_{p} = 0.094$. 
The drop in $\langle V_x\rangle$ is correlated with the formation of
cluster states.
(b) The same for a system with 
$\phi = 0.337$ and $\phi_{p} = 0.02356$, where
the drop in $\langle V_x\rangle$ also
correlates with the onset of clustering.
}
\label{fig:3}
\end{figure}

To show how the onset of clustering correlates with 
the decrease in transport, in
Fig.~3(a) we plot $\langle V_{x}\rangle$ and 
$C_l/N$, the fraction of particles in the largest cluster, 
versus $R_{l}$ for a system with
$\phi = 0.667$ and $\phi_{p} = 0.094$. 
The decrease in $\langle V_{x}\rangle$ coincides
with an increase in $C_{l}/N$, indicating the formation of clusters. 
When $R_l$ is small,
all of the mobile particles are in the gas phase, and if a particle
encounters an obstacle, it has a high probability of quickly changing its
swimming direction and moving away from the obstacle.
As $R_l$ increases and the system enters the cluster phase,
an obstacle can pin
a cluster 
even if
it contacts the obstacle only at a single point, since the cluster has
transient rigidity.
As a result, the moving clusters can become clogged by the obstacles,
reducing the transport.
Since the clusters are dynamic, the particles that belonged to a cluster
become unpinned once the cluster breaks apart. 
The lifetime of individual clusters increases with
increasing $R_{l}$ and the longer-lived clusters 
remain clogged for longer periods of time, resulting in
a gradual reduction in $\langle V_{x}\rangle$ with increasing $R_{l}$ as
shown in Fig.~2. 
For $\phi_{p} > 0.0785$, in the cluster regime 
where  $\langle V_{x}\rangle$ decreases with increasing
$R_l$ we find
$V_{x}(R_{L}) \propto R^{\alpha}_{L}$, with $\alpha = 1.31 \pm 0.02$ \cite{S}. 
In Fig.~3(b) we plot $C_l/N$ and $\langle V_{x}\rangle$ vs 
$R_{l}$ for a system with $\phi = 0.337$ and $\phi_{p} = 0.02356$.
The transport is almost constant at the values of $R_{l}$ 
where clustering is absent; however, at the onset of clustering
where $C_{l}/N$ increases,
$\langle V_{x}\rangle$ begins to drop rapidly, showing the correlation 
between the clustering and the drop in the transport.         

\begin{figure}
\includegraphics[width=3.5in]{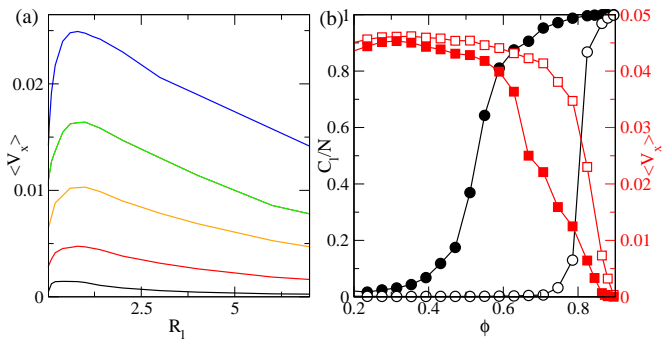}
\caption{
(a) $\langle V_{x}\rangle$ vs $R_{l}$ for 
systems with $\phi_{p} = 0.0942$ 
at $\phi = 0.667$, 0.746, 0.785, 0.8246, and
$0.864$ (from top to bottom) 
showing the decrease in the transport for increasing $\phi$. 
(b) $C_l/N$ (circles) and $\langle V_{x}\rangle$ (squares) 
vs $\phi$ for samples with $\phi_{p} = 0.0235$ at 
$R_{l} = 0.04$ (open symbols) and $R_{l} = 20$ (filled symbols) 
showing the correlation between the
onset of clustering and the drop in the transport.    
}
\label{fig:4}
\end{figure}

We next show how the transport is affected by the particle 
density and jamming effects.   
The jamming concept was originally applied to non-active particles 
in the zero-temperature limit, where 
it was shown that as a function of increasing density the system 
becomes jammed at $\phi_j$, termed point J \cite{33}. 
In 2D assemblies of bidisperse disks,
point J falls at $\phi_j = 0.844$ \cite{34}, while
for monodisperse disk assemblies the system 
forms a hexagonal solid at $\phi = 0.9$. 
Recently it was shown that the addition of quenched disorder to a jamming
system decreases the jamming density,
where jamming was defined to occur when the drift velocity
$\langle V_{x}\rangle = 0$ under an external applied drive \cite{35}.   
In the active matter system for the non-active limit of $R_{l} = 0$, we
can obtain a jammed state with $\langle V_x\rangle=0$ by increasing $\phi$ for
fixed $\phi_p$ or increasing $\phi_p$ for fixed $\phi$.
In Fig.~4(a) we plot $\langle V_{x}\rangle$ vs 
$R_{l}$ for samples with $\phi_{p} = 0.0942$ at 
$\phi = 0.667$, 0.746, 0.785, 0.8246, and $0.864$. 
Each curve has an a optimal $R_{l}$ for transport, and
the magnitude of $\langle V_x\rangle$
{\it decreases} with increasing $\phi$. 
For $\phi > 0.746$, $\langle V_{x}\rangle = 0$ at $R_{l} = 0$, 
corresponding to the
jammed phase. 
For nonzero $R_l$,
the decrease in $\langle V_{x}\rangle$ with increasing $\phi$ is 
due to the formation of clusters which increase in size 
with increasing $\phi$.     
For moderate to large values of $\phi_p$, we find that
the system reaches the $\langle V_x\rangle=0$ jammed state
if infinite run times are used, so that complete jamming can occur
at the limits $R_l=0$ and $R_l=\infty$.
In Fig.~4(b) we plot $C_l/N$ and $\langle V_x\rangle$
versus $\phi$ for samples with $\phi_p=0.0235$ at
$R_{l} = 0.04$ and $R_{l} = 20$.
For $R_{l} = 0.04$,  $\langle V_{x}\rangle$ is constant over most
of the range of $\phi$
but rapidly decreases to a value near
zero at $\phi = 0.9$, where we also find a rapid increase in $C_l/N$.
For $R_{l} = 20$, the onset of clustering occurs at a lower
value of $\phi$ and is also correlated with a
drop in $\langle V_{x}\rangle$.
Our results suggest the possibility of adding a new axis to 
the jamming phase diagram \cite{33} corresponding to the activity
$R_l$, where
for large activity there could be another critical jamming point $A_{j}$ similar
to point J.  

We find that increasing the run length or particle density can 
substantially decrease the mobility
of active matter particles due to cluster formation and local clogging. 
In order to avoid
such clogging effects, new types of motion rules could be devised
for the active particles, such as having the particles
reverse or randomize their swimming direction if they become immobilized by
an obstacle or by entrapment in a cluster.
It would interesting to explore whether
biologically relevant active matter such as swimming organisms 
have evolved methods to avoid self-clogging. 
Other types of behavior could arise for different types of obstacles;
for example, instead of impenetrable obstacles, the disorder could take the
form of sticky sites that reduce the mobility of the particles but that
permit their passage.
Such pointlike pinning sites would be effective at pinning individual
particles, but would have difficulty trapping an extended object such as
a cluster.

{\it Summary--}  
We have examined the transport of run-and-tumble  active matter particles
driven through quenched disordered environments. 
Unlike thermalized
particles driven over random disorder, where 
previous studies have shown that the transport is generally increased
when the thermal fluctuations increase,
we find that for active matter the transport first increases and
then decreases as the activity level or running length increases
due to the formation
of living crystals that can be locally jammed or clogged. 
The system becomes effectively jammed at lower densities as the
run length increases,
suggesting that activity
could form a new axis of the jamming phase diagram. 
Since recent theoretical work has shown that run-and-tumble dynamics
and active Brownian motion can be mapped to each other \cite{33}, 
our results can be generalized to other
active systems driven over quenched disorder.

\acknowledgments
We thank L. Lopatina for useful discussions. This work was carried out under the auspices of the 
NNSA of the 
U.S. DoE
at 
LANL
under Contract No.
DE-AC52-06NA25396.

\vfill
\eject
\input{EPAPS.tex}

\end{document}

%% file: EPAPS.tex

\section{SUPPLEMENTARY MATERIAL
for Active Matter Transport and Jamming on Disordered Landscapes
}

\setcounter{figure}{0}
\begin{figure}[b]
\includegraphics[width=3.4in]{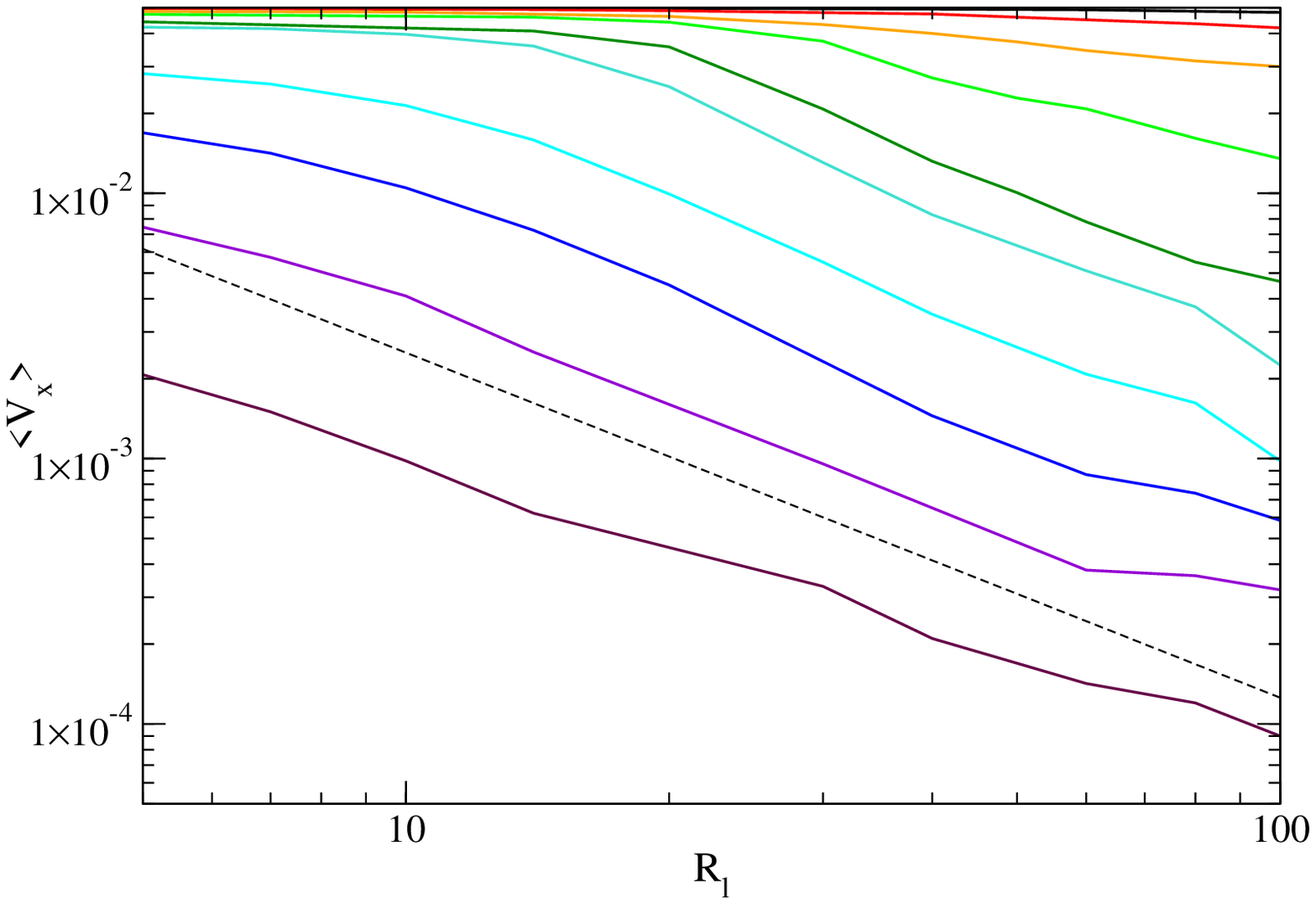}
\caption{
A log-log plot of $\langle V_{x}\rangle$ vs $R_{l}$ 
for systems with $\phi = 0.667$ and $F_{d} = 0.05$ at 
$\phi_{p} = 0.00039$, 0.00157, 0.00472, 0.0785, 0.0157, 0.02356, 0.055, 
0.0942, 0.01413, and $0.188$, from top to bottom.
The dashed line is a fit to a power law with 
$V_{x}(R_{l}) \propto R^{\alpha}_{l}$, with $\alpha = 1.31 \pm 0.02$. }
\end{figure}
\begin{figure}[t]
\includegraphics[width=3.4in]{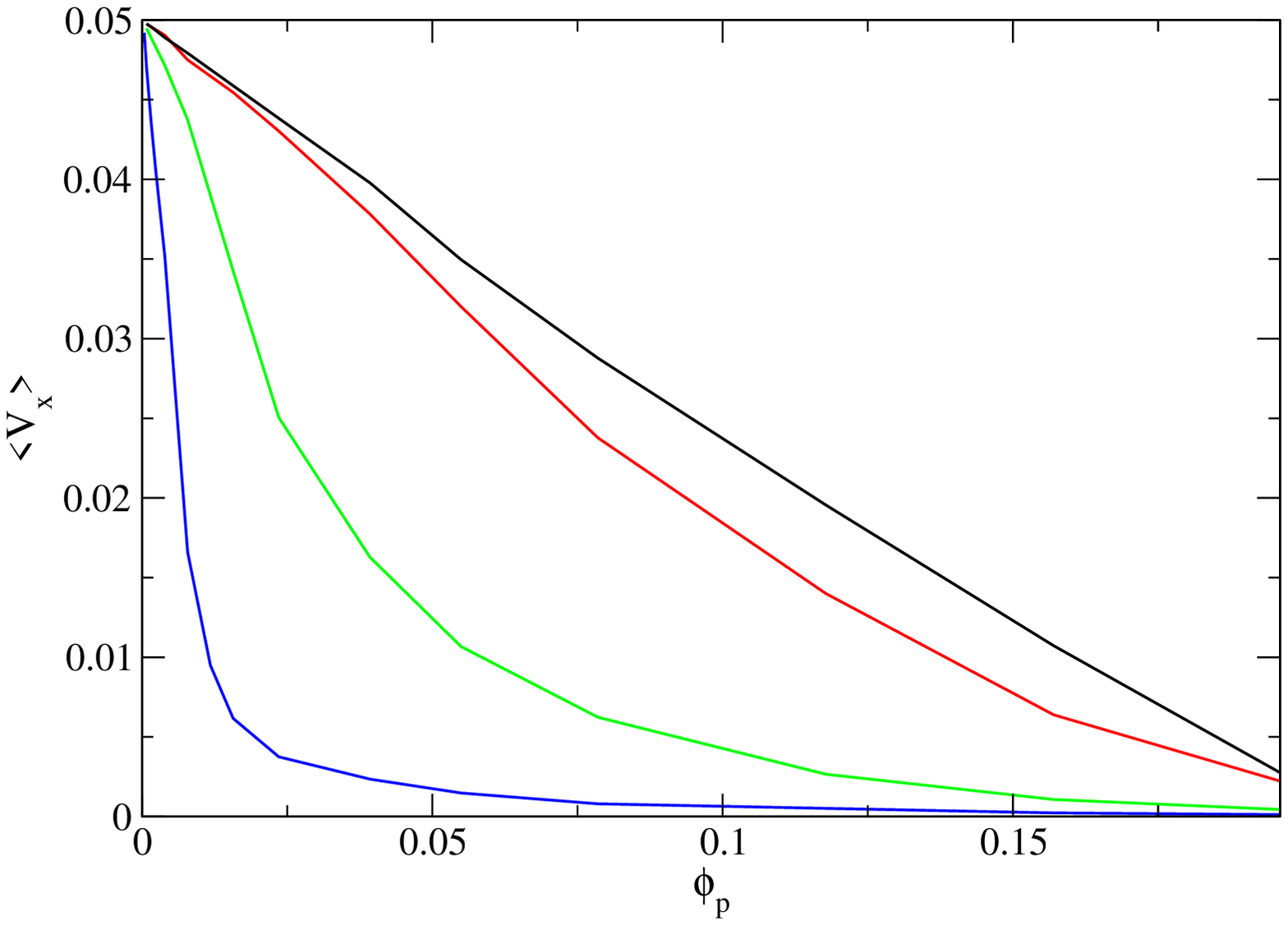}
\caption{
$\langle V_{x}\rangle$ vs $\phi_{p}$ at a fixed $\phi = 0.67$ 
for $R_{l} = 1.0$, 4.0, 20.0, and $80.0$, from top to bottom.
At small run lengths, 
$\langle V_{x}\rangle$ decreases linearly with increasing obstacle 
density, while for large
run lengths the decrease is much more rapid and approximately follows
$\langle V_{x}\rangle \propto 1/\phi_{p}$.   }
\end{figure}
